\documentclass[]{article}
\usepackage{multicol}
\usepackage{apj}

\newcommand {\h} {$h^{-1} \, Mpc \,$}
\newcommand {\kk} {$h^{-1} \, kpc \,$}
\newcommand {\ks} {$km~s^{-1} \;$}

\def\lesssim{\mathrel{\hbox{\rlap{\hbox{\lower4pt\hbox{$\sim$}}}\hbox{$<$}}}}
\def\gtrsim{\mathrel{\hbox{\rlap{\hbox{\lower4pt\hbox{$\sim$}}}\hbox{$>$}}}}

\begin{document}


\vspace{15mm}
\begin{center}
\uppercase{Optical Mass Estimates of Galaxy Clusters}\\   
\vspace*{1.5ex}
{\sc Marisa Girardi$^{1,2,3}$, Giuliano Giuricin$^{2,3}$,
Fabio Mardirossian$^{1,2,3}$,
Marino Mezzetti$^{2,3}$, \\ and Walter Boschin$^{2,3}$}\\
\vspace*{1.ex}
{\small
$^1$Osservatorio Astronomico di Trieste, Via Tiepolo 11, I-34131 
Trieste, Italy\\ $^2$SISSA, via Beirut 4, 34014 Trieste, Italy\\
$^3$Dipartimento di Astronomia, Universit\`{a} degli Studi di Trieste, 
Trieste, Italy\\ E-mail: girardi, giuricin, mardiros, mezzetti,  @sissa.it;
boschin@newton.sissa.it}
\end{center}
\vspace*{-6pt}

\begin{abstract}

We evaluate in a homogeneous way the optical masses of 170 nearby
clusters ($z\le 0.15$).  The sample includes both data from the
literature and the new ENACS data (Katgert et al. 1996, 1998).

On the assumption that mass follows the galaxy distribution, we 
compute the masses of each cluster by applying the virial theorem to the
member galaxies. We constrain the masses of very substructured
clusters (about $10\%$ of our clusters) between two limiting values.

After appropriate rescaling to the X-ray radii, we compare our optical
mass estimates to those derived from X-ray analyses, which we compiled
from the literature (for 66 clusters).  We find a good overall
agreement.  This agreement is expected in the framework of two common
assumptions: that mass follows the galaxy distribution, and that
clusters are not far from a situation of dynamical equilibrium with
both gas and galaxies reflecting the same underlying mass
distribution.

We stress that our study strongly supports the reliability of present
cluster mass estimates derived from X-ray analyses and/or
(appropriate) optical analyses.


\vspace*{6pt}
\noindent
{\em Subject headings: }
galaxies: clusters: general - galaxies: distances
and redshifts - X-rays: galaxies - cosmology: observations.
\end{abstract}

\begin{multicols}{2}

\section{INTRODUCTION}

The knowledge of the properties of galaxy clusters plays an important
role in the study of large scale structure formation.  In particular,
the observational distribution of the abundance of galaxy clusters as a
function of their mass places a strong constraint on cosmological
models (e.g., Bahcall \& Cen 1993; Borgani et al. 1997; Gross et al. 1998;
 White, Efstathiou \& Frenk 1993).  Moreover, recent studies stress the
need for  having reliable estimates of cluster masses to constrain the
ratio between the baryonic to total mass and the consequent value of
$\Omega_0$ (e.g., White \& Frenk 1991; White et al. 1993b).

Indeed, the estimate of cluster masses is not an easy task, in spite of
the various methods which are available.  The application of the
virial theorem to positions and velocities of cluster member galaxies
is the oldest method of cluster mass determination (e.g., Zwicky
1933).  More recent methods are based on the dynamical analysis of hot
X-ray emitting gas (e.g., Cowie, Henriksen, \& Mushotzky 1987; Eyles
et al. 1991) and on gravitational lensing of background galaxies (e.g
Grossman \& Narayan 1989).

Mass estimates derived from the dynamical analysis of gas or member
galaxies which are based on the Jeans equation or its derivations,
such as the virial theorem, assume that clusters are systems in
dynamical equilibrium (e.g., Binney \& Tremaine 1987).  This
assumption is not strictly valid; in fact, although clusters are bound
galaxy systems, they have collapsed very recently or are just now
collapsing, as is suggested by the frequent presence of substructures
(e.g., West 1994).  However, some analyses suggest that the estimate
of optical virial mass is robust against the presence of small substructures
(Escalera et al. 1994; Girardi et al. 1997a; see also Bird 1995 for a
partially different result), although it is affected by strong
substructures (e.g., Pinkney et al. 1996).  Similar results come from
studies based on numerical simulation (e.g., Schindler 1996a; Evrard,
Metzler, \& Navarro 1996; Roettiger, Burns, \& Loken 1996) for X-ray
masses estimated with the standard $\beta$-model approach (Cavaliere \& Fusco
Femiano 1976), although some authors have claimed there is a systematic mass
underestimation (e.g.; Bartelmann \& Steinmetz 1996).

Dynamical analyses based on galaxies have the further drawback that the
mass distribution or (alternatively) the velocity anisotropy of galaxy
orbits should be known a priori.  Unfortunately, the two quantities
cannot be disentangled in the analysis of the observed velocity
dispersion profile, but only in the analysis of the whole velocity
distribution which, however, requires a large number of galaxies (of
the order of several hundreds; e.g. Dejonghe 1987;  
Merritt 1988; Merritt \& Gebhardt
1994). Without some information from the relative distribution of
dark and galaxy components, the virial theorem places only
order-of-magnitude constraints on the total mass (e.g., Merritt
1987). The usual approach is to apply the virial theorem by
assuming that mass is distributed 
like the observed galaxies 
(e.g., Giuricin, Mardirossian, and Mezzetti 1982; Biviano et
al. 1993).  This assumption is supported by several  pieces of evidence
coming both from optical (e.g., Carlberg, Yee, \& Ellingson 1997) and
X-ray data (e.g., Watt et al. 1992; Durret et al. 1994;
Cirimele, Nesci, \& Trevese 1997), as well as from gravitational
lensing data, which, however, suggest a smaller core radius (e.g., Narayan \&
Bartelmann 1997).

The mass estimates derived from gravitational lensing phenomena are
completely independent of the cluster dynamical status, but a good
knowledge of cluster geometry is required in order to go from the
projected mass to the cluster mass (e.g., Fort 1994).  Moreover, strong
lensing observations give values for the mass contained within very
small cluster regions ($\lesssim$ one hundred of kpc) and weak lensing
observations are generally more reliable in providing the shape of the
internal mass distribution rather than the amount
of mass (e.g., Squires \& Kaiser 1996).

Up to now, few studies have dealt with wide comparisons between mass
estimates obtained by different methods for the same cluster.  Wu \&
Fang (1996; 1997) found that masses derived from gravitational lensing
analyses are higher than those from X-ray analyses by a factor of
2, but agree with those from galaxy analyses.  Indeed, mass
estimates from lensing seem to agree with X-ray estimates when
clusters are relaxed (e.g., Allen 1997).  However, Wu \& Fang's works
concern only clusters which lie at moderate redshifts and show
gravitational lensing phenomena which could be enhanced in the
presence of substructures (e.g., Miralda-Escud\'e 1993; Bartelmann,
Steinmetz, \& Weiss 1995).  For nearby clusters, there is a trend to
obtain larger masses from galaxy analyses than from X-ray
analyses (e.g., Cowie et al. 1987; Mushotzky et al. 1995; David,
Jones, \& Forman 1995), but acceptable agreement exists in some
individual cases (e.g., for the Coma cluster, Watt et al. 1992).

The classical approach of the virial theorem based on measurements of
discrete velocities bears re-examining owing to the large new data sets
which are now becoming available for nearby clusters, i.e. the ESO Nearby
Abell Clusters Survey (ENACS) by Katgert et al. (1996, 1998).
Moreover, the fair level of consistency among recent estimates of velocity
dispersion of member galaxies resulting from different membership
assignment procedures (cf. Fadda et al. 1996, hereafter F96, and
Mazure et al. 1996) makes us confident of the robustness of our
approach.

The aim of this work is to obtain reliable mass estimates.  These mass
estimates will be used in the computation of the mass function of
nearby clusters (Girardi et al. 1998).

The paper is organized in the following manner.  We describe the data
sample and our selection procedure for cluster membership assignment
in \S~2. We briefly describe the methods used to compute cluster masses by
using member galaxies in \S~3. By assuming that mass follows the galaxy
distribution, we compute virial mass estimates in \S~4, and we verify
their consistency with the results of the Jeans equation in \S~5.  The
strongly substructured clusters are analyzed in \S~6.  We compare our
mass estimates with those derived from X-ray analyses in \S~7.  We
discuss our results in \S~8.  We give a brief summary of our main
results and draw our conclusions in \S~9.

Unless  otherwise specified,  we give errors
at the 68\% confidence level (hereafter c.l.)

A Hubble constant of 100 $h$ \ks $Mpc^{-1}$ is used throughout.

\section{THE DATA SAMPLE}

Our data sample is essentially an updating of that of F96, which is a
 compilation of data on nearby clusters ($z\le0.15$) coming from the
 literature and from the new ENACS data set (Katgert et al. 1996,
 1998).  In this work we consider the clusters for which F96 gave an
 estimate of velocity dispersion.  Moreover, we consider three new
 clusters (A3880, S639, S721) and we update the original data sets for
 the clusters A576, A1016, A3528N, A3556, and A3733.  Table~1 lists
 the total cluster sample. In Col.~(1) we list the cluster name; in
 Col.~(2) the number of galaxies with measured redshift in each
 cluster field; in Col.~(3) the Abell richness class; and in Col.~(4)
 redshift references.  The Supplementary Clusters of the ACO catalog
 (Abell, Corwin, \& Olowin 1989)  with Abell number counts $N_c<30$ are
 classified as belonging to the richness class $R=-1$.

In order to select member galaxies, we apply the same procedure as
F96 to the new (or updated) clusters. Here we summarize the main points.  We
use the adaptive kernel method (Pisani 1993) to find the
significant peaks in velocity distributions and then the ``shifting
gapper'' to reject galaxies that are too far in velocity (by $\ge 1000$ \ks)
from the main body of galaxies at a given distance from the cluster
center (generally within a shifting annulus of 0.4 \h).

Our  total sample contains 170 clusters, with each cluster having at least
30 galaxies with available redshift in the original field. Ten
clusters consist of two clumps obviously separated in spatial position
(denoted by their spatial orientation, e.g. A548NE) 
or two significant peaks in the velocity distribution (denoted by
"a,b,...", e.g. A151a).  Eighteen clusters
show a secondary peak in the velocity distribution, which strongly
overlaps the primary one so that their separation is
uncertain. Indeed, they could be either two different systems or two parts
of the same substructured cluster and, like F96, we treat them
by considering the peaks either joined or  disjoined.

First we compute masses for 152 clusters which do not show ambiguity
in the dynamics and allow more refined analyses, while the other 18
clusters are analyzed in \S~6.

\section{CLUSTER MASSES BASED ON THE  DYNAMICS OF MEMBER GALAXIES}

The standard methods used to estimate the cluster mass from member galaxies
require that galaxies be in equilibrium within the cluster potential.
The cluster mass is then recovered from the knowledge of positions and
velocities of the same population of galaxies which are taken as
tracers of the cluster potential.  We summarize the main points of the
procedures used to estimate masses from the Jeans equation and the
virial theorem.

\subsection{THE MASS DERIVED FROM THE JEANS EQUATION}

In principle, one can estimate the cluster mass within a radius $r$,
$M_J(<r)$, by using the Jeans equation, coupled with the equation which
links the two observable quantities $\Sigma(R)$ and $\sigma_P(R)$,
i.e.  the projected galaxy number density and the projected velocity
dispersion as a function of the projected radius $R$:

\begin{equation}
\frac{d(\rho \sigma_r(r)^2)}{dr}+\frac{2\rho(r)\beta\sigma_r^2}{r}= -\frac{G\rho(r)M_J(<r)}{r^2}, 
\end{equation}

\begin{equation}
\sigma_P^2(R)\Sigma (R)=2\int^\infty_R 
\rho(r)\sigma_r^2(r)(1-\beta\frac{R^2}{r^2})\frac{r}{\sqrt{r^2-R^2}}dr
\end{equation}

\noindent 
where r is the distance from the cluster center,
$\rho(r)$ is the spatial number density of galaxies linked to
$\Sigma(R)$ via the Abel integral, $\sigma_r(r)$ is the radial
component of velocity dispersion $\sigma(r)$, and
$\beta(r)=1-\sigma_{\theta}^2/\sigma_r^2$ is the velocity anisotropy
parameter (e.g., Binney \&
Tremaine 1987). 

Unfortunately, there are three unknowns ($M(<r)$,$\sigma(r)$,
$\beta(r)$) and only two equations. In order to solve these equations
it is therefore necessary to make some assumptions. It seems natural
to assume knowledge of either $\beta(r)$ or $M(r)$, and then to
evaluate the remaining two functions so that they are consistent to
the observed velocity dispersion profile (e.g., Merritt 1987).

The virial theorem derives from the Jeans equation via an integration step.
It relates the global kinetic energy with the potential one ($2T+U=0$,
e.g.  Binney and Tremaine 1987) and is usually used to compute virial
masses. 

A different derivation (Heisler, Tremaine, \& Bahcall 1985) gives an
alternative mass estimator, the projected mass, which assumes an a
priori velocity anisotropy parameter (see also Perea, del Olmo, \&
Moles 1990 for the non-isotropic case). Since we do not use this
estimator, we do not give any details about it  and refer only 
to the original papers.

\subsection{THE MASS DERIVED FROM THE VIRIAL THEOREM}

The total virial mass of the cluster, $M_V$, depends on the global
velocity dispersion, $\sigma$ and the spatial distribution of the
galaxy population (e.g., Merritt 1988):

\begin{equation}
M_V=\frac{<v^2>}{G<r^{-1} F>},
\end{equation} 

\noindent where the brackets indicate spatial averages over the observed
sample of $N$ galaxies, $r$ are the galaxy distances from the cluster center,
and $v$ are the galaxy velocities referred to the cluster mean velocity.  The
function $F(r)$ is the mass fraction within $r$ and depends on the
(generally) unknown form of mass distribution.

If mass is distributed like the observed galaxies
(i.e., $\rho_{mass}\propto \rho$), then the appropriate
form of eq.~3 is (Limber \& Mathews 1960):

\begin{equation}
M_V=\frac {<v^2>} {G<r_{ij}^{-1}>}=\sigma^2 R_V/G
\end{equation} 

\noindent where $R_V$ is the virial radius
which depends on $r_{ij}$, i.e. the distance between any pair of
galaxies.

From the observational point of view, the large advantage of the virial
theorem is that the global projected velocity dispersion $\sigma_P$ and,
consequently, the total mass are independent of possible anisotropy of
galaxy velocities, always being $\sigma^2=3\sigma
_P^2$ for spherical systems (e.g., The \& White 1986; Merritt 1988). 
Therefore, in the case of spherical systems, 
for the respective projected quantities $\sigma_P$ and $R_{PV}$,
eq.~4 becomes:

\begin{equation}
M_V=3\pi/2 \cdot \frac {<V^2>}{G<R_{ij}^{-1}>}=3\pi/2 \cdot \sigma_P^2 R_{PV}/G. 
\end{equation}

The uniform weighting of galaxies in
computing spatial averages is largely assumed in the literature
and it is justified by the absence of pronounced luminosity
segregation in galaxy clusters (e.g., Adami, Biviano, and Mazure
1998; Biviano et al. 1992; Stein 1997).
In this framework, the projected velocity dispersion, $\sigma_{P}$,
and the projected virial radius, $R_{PV}$,
are estimated as: 

\begin{equation}
\sigma_P=\sqrt{(\Sigma_i V^2/(N-1))},
\end{equation}

\begin{equation}
R_{PV}=N(N-1)/(\Sigma_{i\ne j} R_{ij}^{-1}).
\end{equation}

When  the system is not entirely included in the observational sample,
as is usual in
galaxy clusters, the usual formula of the virial theorem
$2T+U=0$ should be replaced by $2T+U=3PV$ where $3PV$ is the 
surface term (e.g. Binney \& Tremaine 1987;
Carlberg et al. 1996; Carlberg et al. 1997a; The \& White 1976).
Therefore, a correction, $C$, should be applied
to the usual formula of virial mass which, otherwise, overestimates
the true cluster mass.  In particular, if mass follows the galaxy
distribution, the corrected virial mass, $M_{CV}$ is:

\begin{equation}
M_{CV}=M_V-C=M_V(1-4 \pi b^3 \frac{\rho(b)}{\int^b_0 4\pi r^2\rho dr} (\sigma_r(b)/\sigma(<b))^2),
\end{equation}

where $\sigma(<b)$ refers to the integrated velocity dispersion within 
the boundary radius $b$.

\section{ESTIMATING OPTICAL CLUSTER MASSES}

In this section we compute the cluster masses on the hypothesis
that galaxy number-distribution traces mass
distribution. Therefore, the results of \S~4.5 and 4.6 depend on this
assumption.

From the observational point of view, compared to use of the Jeans
equation, the virial theorem has the great advantage of using the more
robust integrated values of $\sigma_P$ rather than the differential
values.  Since the computation of the observational dispersion profile
$\sigma_P(R)$ requires a large number of galaxies, we can compute it
only by combining together the data of many clusters, without
preserving cluster individuality.  Therefore, we use the virial
theorem to compute the mass of each individual cluster and then we use
the Jeans equation to check our results on ``average'' clusters
(\S~5).

\subsection{VELOCITY DISPERSION}

The problems concerning the estimate of the velocity dispersion,
$\sigma_P$, have been faced in two previous papers (F96; Girardi et
al. 1996). These studies found that the observed profiles 
$\sigma _P(<R)$, where the velocity dispersion is computed by including
larger and larger cluster regions, 
may show strongly increasing (or
decreasing) behaviors in the central cluster regions, but 
flatten out in the external regions, suggesting that in these
regions the (almost constant) value of $\sigma_P$ is no longer
affected by possible velocity anisotropies.  Therefore, in eq.~5 one
can consider as physically meaningful only the value of $\sigma_P$
computed by considering all galaxies within a large radius.

We use the values of projected velocity dispersions as tabulated in
Table~3 of F96 and, for new (and updated) clusters, we adopt their
same procedure to estimate $\sigma_P$.  In particular, we use robust
estimates (computed via the ROSTAT routines by Beers, Flynn, \&
Gebhardt 1990), obtained by applying the relativistic correction and
the usual correction for velocity errors (Danese, De Zotti, \& di
Tullio 1980).

\subsection{THE RADIUS OF VIRIALIZATION}

Since the application of the virial theorem is really meaningful only
when the system is in dynamical equlibrium within the region
considered, the natural choice is to compute cluster masses within the
radius of virialization, $R_{vir}$.  The precise computation of the
radius of virialization is possible only after the computation of
cluster mass (\S4.6) and would require a recursive procedure. This
will be given in Girardi et al. (1998) for a few standard cosmological
scenarios.  Here, we give a reasonable a priori approximation.

In the framework of the theory of a spherical model for nonlinear
collapse, the relation between the present density of a collapsed
(virialized) region and the present cosmological density is
$\rho_{vir}(t_0) = 18\pi^2 \rho_0= 18 \pi^2 \times 3 H_0^2/8\pi G$
(for a $\Omega_0=1$ Universe).  As a first approximation, the virial
mass ($M_{vir}=4\pi R_{vir}^3 \rho_{vir} /3$) can be estimated as
$3\pi /2  \times \sigma_P^2 R_{PV}/G$ (eq.~5).  Therefore:

\begin{equation}
R_{vir}^3=\sigma_P R_{PV}/(6\pi H_0^2),
\end{equation}

\noindent 
where the projected virial radius, $R_{PV}$, depends on the galaxy
distribution.

From  observed galaxy distributions,
Girardi et al. 1995 (hereafter G95; see eqs.~A6,A8)
showed that  the typical $R_{PV}$ is related
to the aperture $A$, i.e. the radius of the sampled
region (here equal to $R_{vir}$), as follows:

\begin{equation}
R_{PV}=1.193 A \frac{1+0.032(A/R_c)}{1+0.107(A/R_c)},
\end{equation}

\noindent where the core radius $R_c$ is on average  $=0.17$ \h~(G95).  
Eqs.~9 and 10 imply the following relation between $R_{vir}$ and 
$\sigma_P$:

\begin{equation}
R_{vir}\sim 0.002 \cdot \sigma_P (h^{-1} Mpc),
\end{equation}

\noindent with $\sigma_P$ given in \ks.

In the following analyses we consider only member galaxies within $R_{vir}$.

\subsection{GALAXY DISTRIBUTION}

We examine the galaxy distribution of the 92 well-sampled clusters, i.e.
the clusters sampled with at least ten galaxies up to $R_{vir}$.
Their galaxy distribution are analyzed in a similar way to that
used  by G95,
i.e. by fitting the galaxy surface density of each cluster to a King
distribution with a variable  exponent (hereafter referred to
as a ``King-like'' profile):

\begin{equation}
\Sigma(R)=\frac{\Sigma_0}{ (1+(R/R_c)^2)^{\alpha}}, 
\end{equation}

\noindent where $R_c$ is the core radius and $\alpha$ is the parameter
which describes the galaxy distribution in external regions
($\alpha=1$ corresponds to the classical King distribution).  This
surface density profile corresponds to a galaxy volume-density
$\rho=\rho_0/(1+(r/R_C)^2)^{3\beta_{fit,gal}/2}$, with
$\beta_{fit,gal}=(2\alpha+1)/3$, i.e.  $\rho(r) \propto r^{-3 \beta_{fit,gal}}$
for $r>> R_C$.  We perform
the fit through the Maximum Likelihood
technique, allowing $R_C$ and $\alpha$ to vary from 0.01 to 1 and from
0.5 to 1.5, respectively,
 and verifying that the fitted profiles are not rejected
by a Kolmogorov-Smirnov test.  

In order to explore the typical galaxy distribution, we re-do the fits
by fixing the value of $\alpha$, the less scattered parameter (see
also G95).  The median value of $\alpha$, with the respective $90\%$
c.l. intervals, is $=0.70_{-0.03}^{+0.08}$, corresponding to a
$\beta_{fit,gal}=0.8$, i.e. to a volume galaxy-density $\rho \propto
r^{-2.4}$.  By fixing the value of $\alpha$, we again fit the galaxy
distribution of each cluster, obtaining median values of
$R_c=0.05_{-0.01}^{+0.01}$ \h~($90\%$ c.l. intervals) and a
$R_c/R_{vir}=0.05$.

As a final check, we analyze the effect of galaxy incompleteness in the
external cluster regions, which could mimic a greater value of
$\alpha$.  We consider a subsample of 27 clusters (denoted by an
asterisk in Table~1) from which it is possible to extract
magnitude-complete samples (at least up to the 80\% completeness
level) according to the prescriptions of the authors of redshift data.
After having performed the two-parameter fit, we find a median value
of $\alpha =0.65_{-0.07}^{+0.05}$ ($90\%$ c.l. intervals), in agreement with the value obtained
from the whole cluster sample.

\subsection{THE VIRIAL RADIUS}

The projected virial radius $R_{PV}$ depends on galaxy distribution
and increases with the cluster aperture $A$
within which it is estimated. Here, we are interested in
determining $R_{PV}$ at $A=R_{vir}$.  We cannot use the direct estimate
of $R_{PV}$ (eq.~7) for the clusters that are not sampled at least up to
$R_{vir}$ (these 50 clusters are sampled up to $R_{max}=0.7\times R_{vir}$, 
on average) or have too small a number of galaxies.

Once the values of $R_C$ and $\alpha$ are known, we can use the
alternative estimate based on the galaxy distribution to compute 
$R_{PV}$ at $R_{vir}$ (eq.~A1 of
G95).  We consider three estimates ($R_{PV}(\alpha,R_C)$,
$R_{PV}(\alpha=0.7,R_C)$, and $R_{PV}(\alpha=0.7,R_C=0.05\times
R_{vir})$), 
respectively based on: a) the values of $\alpha$ and $R_C$
resulting from the above two-parameter fit, b) the median value of
$\alpha$ and the value of $R_C$ resulting from the one-parameter fit,
and c) the median values of $\alpha$ and $R_c$. 
 In particular, we
note that for $\alpha=0.7$, the relation between $R_{PV}$ and $R_C$
for a given cluster aperture $A$ (eq.~1 of G95) is well represented by:

\begin{equation}
R_{PV}=1.189 A \frac{1+0.053(A/R_c)}{1+0.117(A/R_c)}.
\end{equation}

We analyze the 92 well-sampled clusters, which are sampled up to
$R_{vir}$, to verify the acceptability of the three alternative
procedures.  Figure~1 (top-left panel) shows that the virial radius
estimated from the individual fitted values of $R_C$ and $\alpha$,
$R_{PV}(\alpha,R_C)$, is a good alternative estimate of the direct
estimate of $R_{PV}$. When one or two parameters are fixed, the
estimates result in larger and larger scatter in the relation with
$R_{PV}$ (see Figure~1, top-right and bottom-left panels,
respectively), supporting the existence of an intrinsic spread of
cluster parameters.

\includegraphics{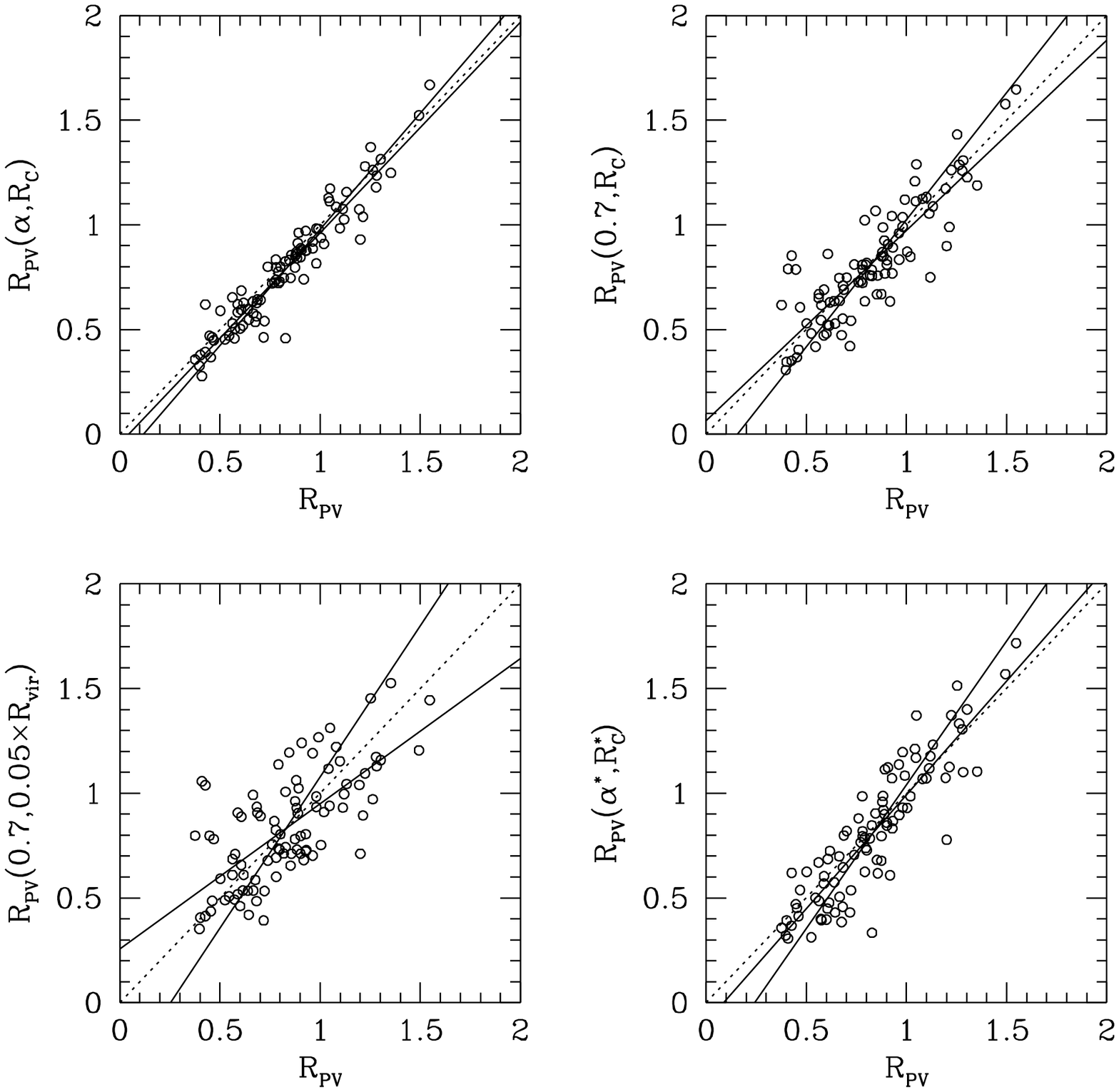}
$\ \ \ \ \ \ $\\
\vspace{8truecm}
$\ \ \ $\\
{\small\parindent=3.5mm {Fig.}~1.---
Four alternative estimates of the virial radius
$R_{PV}$ obtained from galaxy distribution parameters ($\alpha$,
$R_C$) vs. the direct estimate obtained from galaxy separations. The
units are in $h^{-1} Mpc$. The solid lines represent the linear fits
(y vs. x and x vs. y regression lines).
}
\vspace{5mm}

Moreover, in order to check the effect of the limited cluster
aperture, we also compute the estimate $R_{PV}(\alpha^*,R_C^*)$  by
using $R_C$ and $\alpha$ fitted within $0.7 \times R_{vir}$ rather than
within $R_{vir}$. The values of $R_{PV}$ are recovered by a larger scatter
with respect to the optimal case (cf. bottom-right panel with
top-left panel in Figure~1).

Summarizing, we compute $R_{PV}$ (at $R_{vir}$): i) by using the direct
definition of eq.~7 for 100 clusters; ii) by using the estimate
obtained from the fitted values of $\alpha$ and $R_C$ for the 50
clusters with $R_{max}<R_{vir}$; iii) by using the estimate obtained
from the typical values of $\alpha=0.7$ and $R_C=0.05\times R_{vir}$
\h~only for 8 clusters with a very small number of galaxies ($\le 6$).
For case i), i.e. the direct estimate, we compute the jackknife error;
in cases ii) and iii) we adopt the respective s.d. of residuals from
the direct regression line, i.e. 0.15 and 0.19 \h.

The effect of cluster asphericity on the estimate of $R_{PV}$ may be
estimated directly if one assumes that galaxy distributions are
axisymmetric (prolate or oblate), and that isodensity surfaces are
concentric, similar ellipsoids.  We use Monte Carlo simulations to
evaluate the projection factor, $f_R=R_V/R_{PV}$, as a function of the
inclination and of the intrinsic shape.  In particular, we simulate
the galaxy systems by requiring that, along each axis, the particles
fit a 1-D King distribution with a core radius proportional to the
cluster dimension along that axis and we compute a mean value of $f_R$
by averaging 200 simulated systems of 100 particles for each one. Even
if $f_R$ can strongly differ from the spherical symmetric value
(i.e., $\pi/2$), its expectation value is $1.58\pm 0.19$, if we assume
that clusters are prolate and have a Gaussian distribution of intrinsic
axial ratios with a mean of 0.5 and a s.d. of 0.15 (Plionis, Barrow,
\& Frenk 1991).  Moreover, we have verified
that Gaussian or uniform particle distributions give similar mean $f_R$
values to those obtained for a King-distribution.  Owing to large errors
involved in mass computations, we see no reason to adopt a projection
factor different from $\pi/2$.

\subsection{VELOCITY ANISOTROPIES}

In order to compute the $C$-correction  to the virial mass (eq.~8) we
must address the topic of velocity anisotropies of galaxy orbits.

After having assumed the velocity anisotropy parameter $\beta(r)$, one
can compute $\sigma_r(r)$ by solving the Jeans equation (eq.~1) and then
obtain the projected velocity dispersion $\sigma_P(R)$ (eq.~2)
which can be compared with the data.  The
observational $\sigma_P(R)$ is obtained by combining together the
galaxies of many clusters, i.e.  by normalizing distances to $R_{vir}$
and velocities (relative to the mean cluster velocity) to the observed global
velocity dispersion $\sigma_P$.  

The simplest approach is to assume that velocities are isotropic, i.e.
$\beta(r)=0$.  Figure~2 shows the observed profile $\sigma_P(R)$, as
obtained by joining together all clusters, plotted against the theoretical
profiles obtained by using the typical galaxy distribution (\S~4.3).
The model is compatible with our data 
at the $96\%$ c.l..

Den Hartog \& Katgert (1996) and F96  show that
different clusters have markedly different velocity dispersion
profiles. This suggests that the flatness of combined profile is
due to the mixing of clusters with different kinds of velocity
anisotropies and, therefore, with different $C$-corrections.  In
order to give more realistic corrections for individual clusters, we
decide to divide clusters according to the shape of their profiles. 

 We take as a profile indicator, $I_p$, which is the ratio between
$\sigma_P(<0.2\times R_{vir})$, the velocity dispersion computed by
considering the galaxies within the central $0.2\times R_{vir}$, and
the global $\sigma_P$.  The data are good enough to compute $I_p$ for
136 systems.  Note that the individual shape of velocity-dispersion
profiles ($I_p$) does not correlate with cluster velocity dispersion
(see also den Hartog \& Katgert 1996).  We divide clusters in to three
equivalent groups of 45, 46, 45 clusters (hereafter A, B, C clusters,
going from high to low $I_p$, respectively) and build three
observational profiles $\sigma_P(R)$, which are decreasing, flat and
increasing with $R$, respectively.

\includegraphics{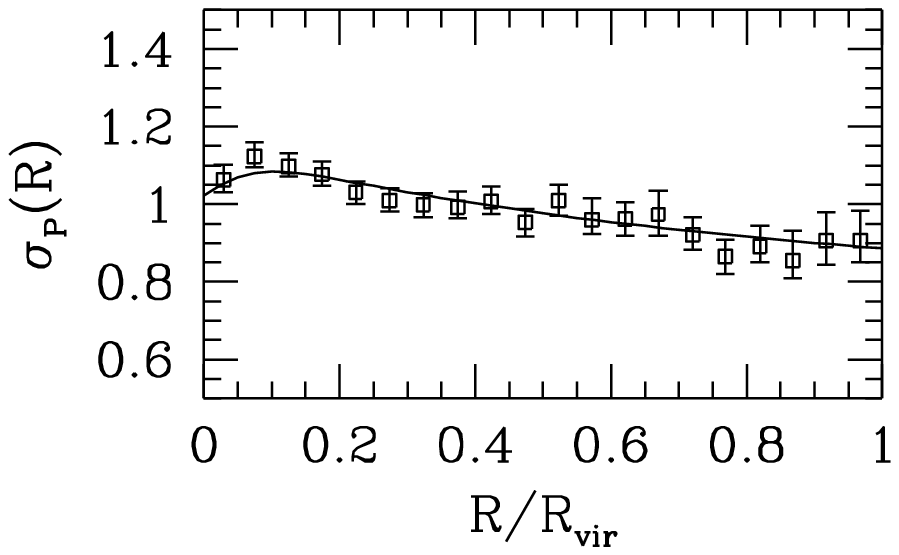}
$\ \ \ \ \ \ $\\
\vspace{4truecm}
$\ \ \ $\\
{\small\parindent=3.5mm {Fig.}~2.---
The (normalized) projected velocity dispersion,
$\sigma_P$, as a function of the (normalized) projected distance from
the cluster center.  The points represent data combined from all
clusters and binned in equispatial intervals. We give the robust
estimates of velocity dispersion and the respective bootstrap
errors. The solid line represents the model for the constant velocity
anisotropy parameter $\beta=0$.
}
\vspace{5mm}

In order to fit the profiles of A and C clusters, we introduce two other
functional forms for $\beta(r)$.  The form $\beta=r^2/(r^2+a^2)$
(e.g., Merritt 1987) describes a cluster with isotropic velocities in
the center and radial velocities in the external regions, as expected
in the case of galaxy infall from the regions around the cluster; this
form gives a decreasing profile (A clusters).  We use the simple form
$\beta=-c/r$ for describing circular velocities in the center and
more isotropic velocities in the external regions, as expected for a very
relaxed cluster undergoing two-body relaxation in the dense central
region; this form gives an increasing profile (C clusters).

We compute theoretical profiles for the cases A, B, C as in the case
of all clusters together.  Figure~3 shows the observed profiles for
the three categories of clusters, each one plotted against the best
theoretical profile. In particular, the fit (acceptable at $\sim 2\%$
s.l.) for A clusters corresponds to $\beta(r)=r^2/(r^2+a^2)$ with
$a=0.1\times R_{vir}$; the fit (acceptable at $\sim 60\%$ s.l.) for B clusters
corresponds to $\beta(r)=0$; the fit (acceptable at $\sim 6\%$ s.l.) for C
clusters corresponds to $\beta(r)=-c/r$ with $c=1\times R_{vir}$.

\includegraphics{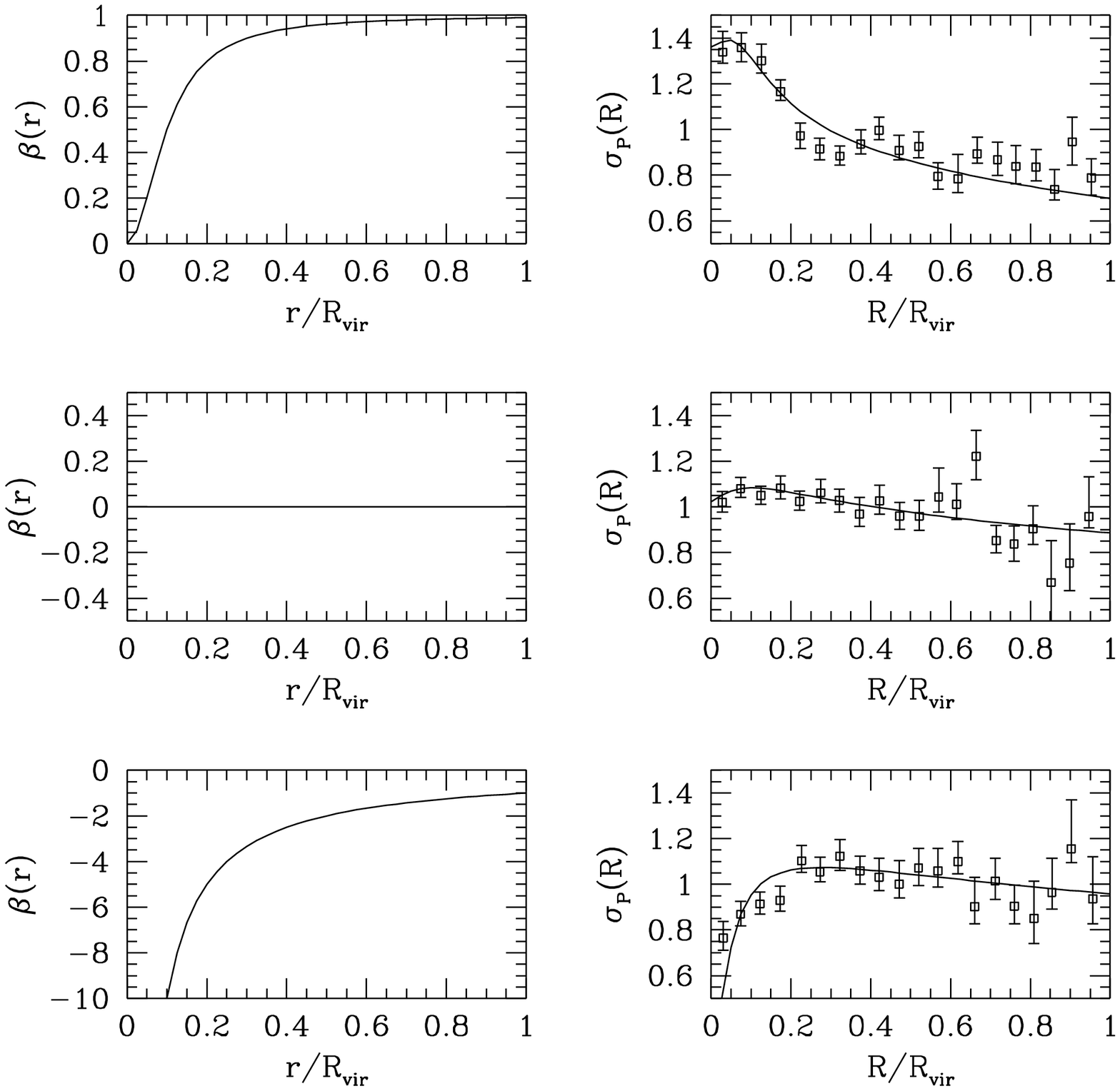}
$\ \ \ \ \ \ $\\
\vspace{8truecm}
$\ \ \ $\\
{\small\parindent=3.5mm {Fig.}~3.---
From top to bottom: we refer to clusters of type
``A'', ``B'', and ``C''.  At right, the (normalized) projected
velocity dispersion, $\sigma_P$, as a function of the (normalized)
projected distance from the cluster center, as in Figure~2. But, here,
the solid line represents the model for the velocity anisotropy
parameter $\beta(r)$ shown in the corresponding panel on the left.
}
\vspace{5mm}

We note that, in this study, which is not devoted to the analysis of
velocity anisotropies, the simple forms adopted for $\beta(r)$ are
adequate for giving a satisfactory estimate of the $C$-correction.
For instance, when we analyze the whole cluster sample by using the
two other functional forms for $\beta(r)$ (instead of 
$\beta(r)=0$) and requiring a compatibility of the fit at only $\sim
2\%$ c.l., we find that the two respective $C$-corrections differ by
$\sim 25\%$, corresponding to a variation for the mass of $\sim 5\%$
(well within the associated observational errors, see \S~4.6). 

\subsection{THE VIRIAL MASS}

In Table~2 we list the cluster parameters computed before: the number
of member galaxies, $N_m$ (Col.~2), contained within the radius
$R_{max}$ (Col.~3) and used to compute the mean (galactocentric)
redshift (Col.~4) and the global projected velocity dispersion
$\sigma_P$ with the respective bootstrap errors (Col.~5); the number
of galaxies, $N$ (Col.~6), within $R_{vir}$ (Col.~7); the values of
$\alpha$ and $R_C$ resulting from the fit to the galaxy distribution
(Cols.~8 and 9, respectively); the projected virial radius, $R_{PV}$,
computed at $R_{vir}$ (Col.~10); the cluster type according to their
velocity dispersion profile, $T$ (Col.~11).

By using $\sigma_P$ and $R_{PV}$, we 
compute the virial mass $M_V$ within $R_{vir}$ through the
standard formula (eq.~5).

However, clusters are well extended outside the virial radius
$R_{vir}$.  By using the theoretical expressions of Gunn \& Gott
(1972) and the same procedure as that described in \S~4.2, we
find that the radius of turn-around and the radius of the region
gravitationally bound to the cluster are $\sim 5-10
R_{vir}$. Therefore, we need to apply the $C$-correction (eq.~8) to
obtain the corrected virial mass $M_{CV}$.  
Here we re-write the $C$-correction of eq.~8 by taking into account the fact
that the boundary radius is $b=R_{vir}$:

\begin{equation}
C=M_V 4\pi R_{vir}^3 \frac{\rho (R_{vir})}{\int_0 ^{Rvir} 4\pi r^2\rho dr} (\sigma_r(R_{vir})/\sigma(<R_{vir}))^2,
\end{equation}

The term $\rho(R_{vir})/\int^{Rvir}_0 4\pi r^2\rho dr$ is computed
for each cluster by using the the galaxy distribution given in eq.~12
with the appropriate individual parameters.

The term $(\sigma_r(R_{vir})/\sigma(<R_{vir}))^2$ is connected to the
presence of velocity anisotropies analyzed
in the above subsection.  

For all clusters joined together, in the case of best fit with the
data ($\beta=0$; see Figure~2), we find that
$(\sigma_r(R_{vir})/\sigma(<R_{vir}))^2$ is $\sim 0.3$.  For the A, B, C
clusters with different kinds of profiles (see Figure~3), the
corresponding values of $(\sigma_r(R_{vir})/\sigma(<R_{vir}))^2$ are
$\sim 0.6, 0.3, 0.2$, respectively.  We adopt the above values of
$(\sigma_r(R_{vir})/\sigma(<R_{vir}))^2$ for A, B, C clusters and the
value obtained for the whole sample ($0.3$) for 24 other clusters
which we cannot assign to any specific kind of profile.

The median value of percentage $C$-correction is 19\% and rises to
39\% when we consider clusters with a decreasing profile towards
external regions (clusters of type A).  

In Table~3 we list the virial masses, within $R_{vir}$, before and
after the correction, $M_V$ and $M_{CV}$ (Cols.~2 and 3,
respectively).  We assume that the percentage errors on $M_{CV}$ are
the same as for $M_V$, i.e. neglecting uncertainties on the 
$C$-correction.

In conclusion, we stress the fact that in the computation of the
$(\sigma_r(R_{vir})/\sigma(<R_{vir}))^2$ term, we have implicitly
assumed that the region outside $R_{vir}$ is in dynamical equilibrium,
too.  Since there is no evidence of a drastic change of profiles of
galaxy density and velocity dispersion at $R_{vir}$ (see, e.g., Fig.~6
of Kent \& Gunn 1982; Fig.~2 of F96), we can regard our estimate of
$C$-correction as a good first approximation.  The analysis of \S~5
will show that our estimate is qualitatively and quantitatively
acceptable.

\section{VERIFYING THE SELF-CONSISTENCY  OF MASS ESTIMATES}

The virial theorem and the Jeans equation should give consistent
mass estimates. Owing to the typically limited number of
member galaxies for each individual cluster, the Jeans equation is not
useful for constraining individual cluster masses, but can be useful for
checking average results.

As a first approximation, cluster galaxies can be described, on
average, by an isothermal distribution, i.e by having a constant
$\sigma$-profile, and isotropic velocities (see \S~4.5). Consequently,
by using the Jeans equation, one expects that the cluster mass
contained within a large radius $r$ is $M_{iso}(<r)=3\beta_{fit,gal}
\sigma_P^2 r/G$ (here $\beta_{fit,gal}=0.8$; see \S~4.3).  We compute
$M_{iso}(<R_{vir})$ for each cluster.  In Figure~4 we show
$M_{iso}(<R_{vir})$ against $M_V$ and $M_{CV}$, respectively. We find
that $M_{iso}$ values are smaller than $M_V$, the virial masses before
the $C$-correction, but are roughly consistent with $M_{CV}$.

\includegraphics{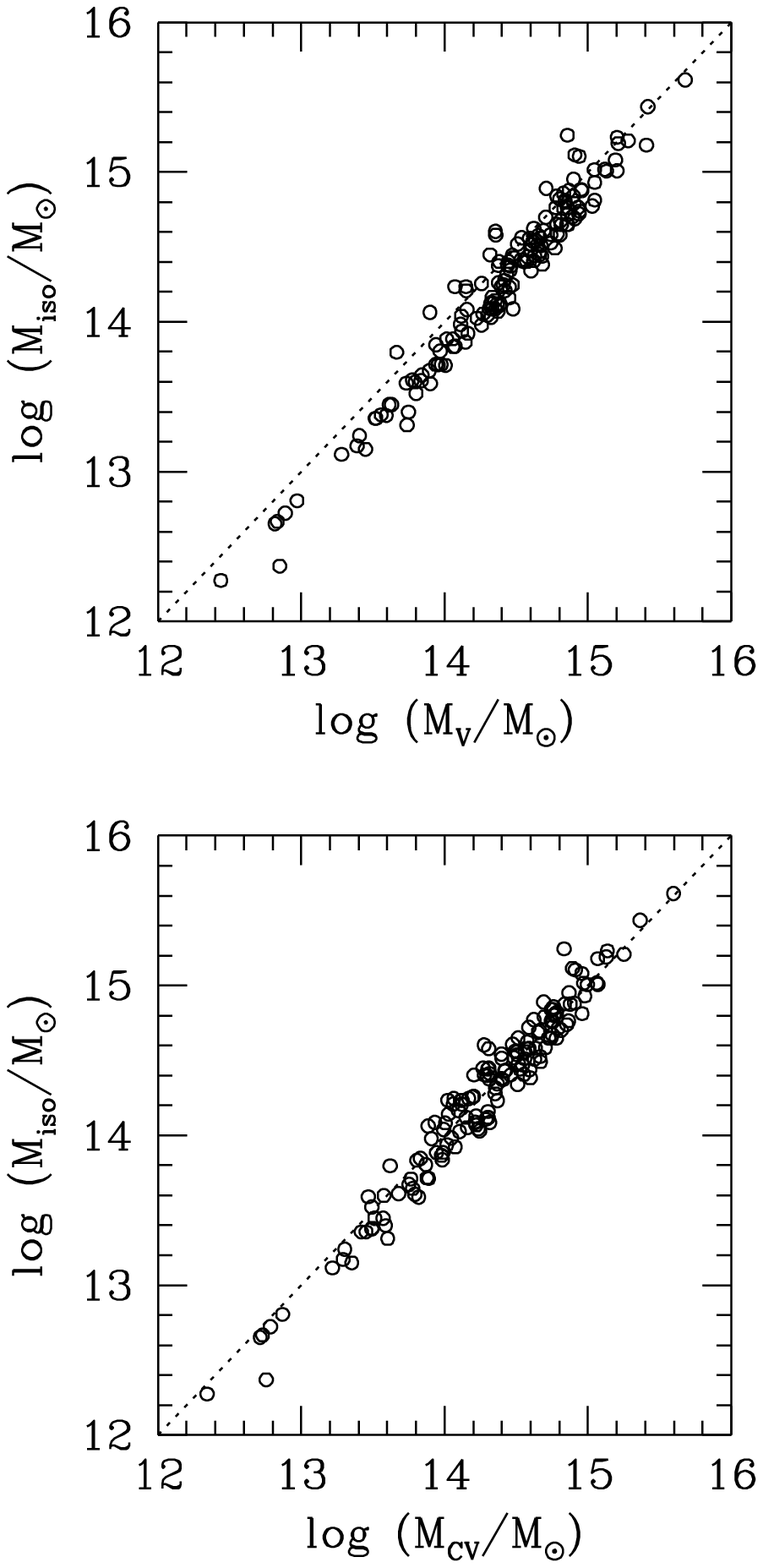}
$\ \ \ \ \ \ $\\
\vspace{8truecm}
$\ \ \ $\\
{\small\parindent=3.5mm {Fig.}~4.---
The comparison between the mass obtained from the
Jeans equation by assuming a constant $\sigma$-profile and isotropic
velocities vs. the virial mass before (upper) and after (lower) the
$C$ correction.  
}
\vspace{5mm}

In order to explore the results for clusters with different velocity
anisotropies, we also consider another approach.  Retaining the
hypothesis that mass follows the galaxy distribution, one can obtain
cluster mass estimates from the Jeans equation, $M_J$, without any
further hypothesis (\S~3.1).  Indeed, by applying the same procedure
as in \S~4.5, but constraining $\sigma_P(<R_{vir})$ to be equal to the
typical (median) observational value, we obtain a value
for the normalization of mass.  In Table~4 we give for each type of
clusters (A, B, C): in Col.~(2) the number of clusters belonging to the
corresponding type; in Cols. (3) and (4) the value of observed
$\sigma_P$ and the corresponding $log(M_J(<R_{vir}))$ computed from
the Jeans equation; in Cols.~(5) and (6) the average values of
$log(M_V)$ and $log(M_{CV})$ as computed in the above subsection.
Again the value of $M_J$ turns out to be smaller than $M_V$, but in
fair agreement with $M_{CV}$.

The general result of our check is that a correction to the standard
virial masses, $M_{V}$, is really needed and that the $C$-correction
appears to be a good estimate of this correction.

\section{STRONGLY SUBSTRUCTURED CLUSTERS}

The strongly overlapping peaks shown by 18 of our clusters in their
velocity distribution may have several explanations. The peaks may be
different systems superimposed along the line of sight, maybe before a
merging, or could indicate the presence of substructures in a single
system.  Remarkably, merging galaxy clumps can also survive the first
encounter (McGlynn \& Fabian 1984; Roettiger, Stone, \& Mushotzky 1998), 
making it more difficult to
understand their real dynamical state. Indeed, the understanding of
the dynamics of these clusters requires individual analysis, with many
more data for each cluster.

Here we limit our analysis to constraining the masses of these 18
clusters between two limiting values, obtained by considering the
peaks to be either joined or disjoined.  We compute these masses by applying
the same procedure as that used above.  The respective cluster parameters and
cluster masses are listed in Tables~5,~6, and ~7,~8 which have the
same format as Tables~2 and ~3, respectively.

The mass estimates strongly depend on the treatment we choose. We find that
the masses obtained in the case of joined peaks 
differ from those obtained in the case of the 
disjoined peaks (by adding the values of the individual peaks)
by a factor of three, on average.

\section{THE COMPARISON WITH X-RAY MASSES}

From the literature, we collect 117 values of X-ray masses, $M_X$,
for 66 clusters.  When a reference source presents
the value of $M_X$ at several radii, we choose the value 
corresponding to  the
largest radius, avoiding possible extrapolation outside the observed
X-ray region.  We avoid strongly substructured clusters, for which the
X-ray analysis may also be problematic (e.g., Roettiger et al. 1996). This
collection of $M_X$ is quite inhomogeneous as regards the method of
analysis, source of X-ray data, cluster richness, and observational
aperture.

In Table~9 we list: the source of the X-ray masses (Col.~2); the value of
X-ray radius (Col.~3), $R_X$, within which $M_X$ (Col.~4) is computed;
our optical masses $M_{opt}$ (Col.~5) obtained by rescaling each $M_{CV}$ to
$R_X$ by using the individual galaxy distribution.  When
the authors did not give errors for $M_X$, we adopt the percentage error
on X-ray temperature $T$, if available, or, otherwise, a bona fide
error of 50\%. For $M_{opt}$ we adopt the same percentage errors 
as for $M_{CV}$.

In Figure~5 we show the comparison between $M_X$ and $M_{opt}$.
We distinguish the data by White, Jones \& Forman (1997), which represent a
large subsample characterized by homogeneity in
the method of analysis and the source of X-ray data.

The masses of individual clusters do not agree in several cases.  The
quality and amount of difference depend not only on the cluster
examined, but also on the reference sources of $M_X$ (e.g., for the
Coma cluster), as expected in an inhomogeneous sample.  The two points
which show the largest discrepancies refer to AWM4 cluster which,
although defined as a poor cluster, shows a very high X-ray
temperature of 3.7 keV (e.g., David et al. 1993 corresponding to
$\sigma_P\sim 800$ \ks~ on the hypothesis of density energy
equipartition between gas and galaxies) when compared with the
observed velocity dispersion (about 100 km/s). This cluster could be a
case of merging between clumps and, therefore, could be farther from a
situation of dynamical equilibrium; however, present galaxy sampling
is so poor that we cannot be more conclusive.

\includegraphics{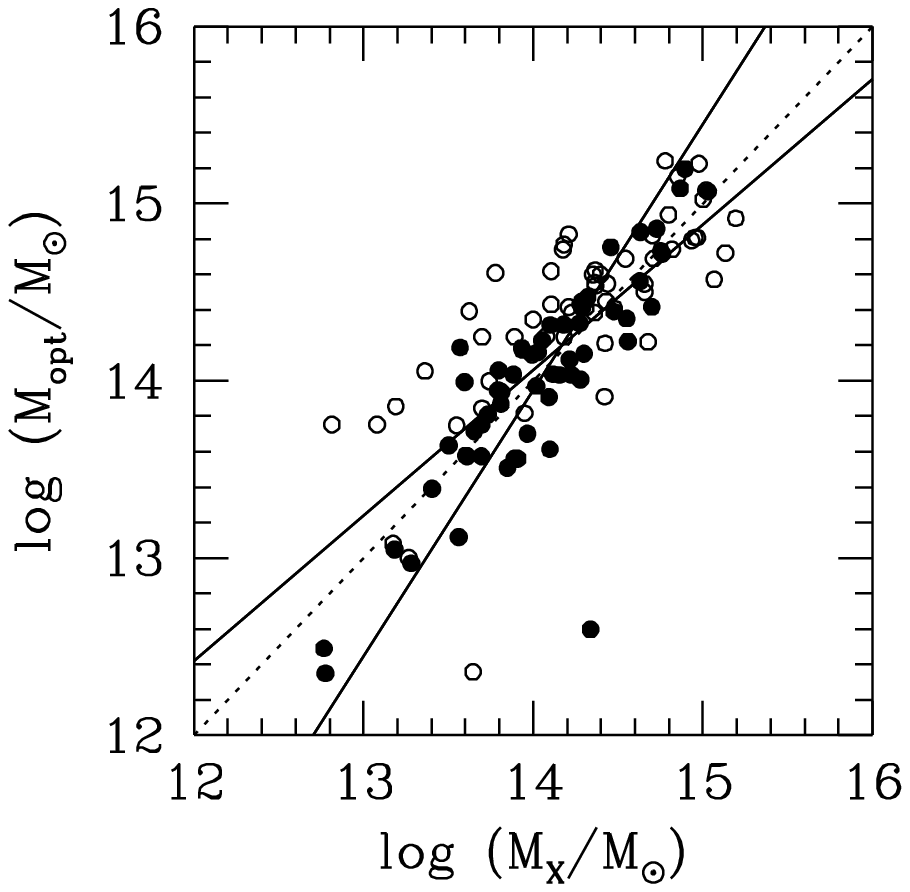}
$\ \ \ \ \ \ $\\
\vspace{6truecm}
$\ \ \ $\\
{\small\parindent=3.5mm {Fig.}~5.---
The comparison between our optical mass estimates
and those derived from X-ray analysis.  The optical masses are
obtained by appropriately rescaling our corrected virial masses to the
same radius as that the X-ray masses refer to.  The solid lines
represent the linear fits (y vs. x and x vs. y regression lines).  The
solid points represent the data by White et al. (1997).
}
\vspace{5mm}

The overall agreement is good.  The bisecting regression line and the
weighted regression line (Press et al. 1982) are
$M_{opt}=10^{(-1\pm1)}\times M_X^{(1.11\pm0.07)}$ and
$M_{opt}=10^{(0.4\pm0.4)}\times M_X^{(1.01\pm0.03)}$, respectively.
The s.d.  of residuals from the weighted regression line gives a
typical deviation of mass values of $\sim 30\%$.  Part of this scatter
is probably due to the inhomogeneity of the X-ray data compilation. In
fact, by considering only the data by White et al. (1997) we reduce
the deviation of mass values to $\sim 25\%$.  According to the
$\chi^2$ fit probability, the observed scatter cannot be explained
only by the errors on masses unless the percentage errors are
underestimated by a factor of two.

\section{DISCUSSION}

We find a fair average agreement between our optical masses and X-ray
masses.  

The previous tendency of finding optical estimates larger than X-ray
estimates (Mushotzky et al. 1995; David et al. 1995) can be explained
by the fact that our mass estimates are generally smaller than the
previous ones.  After appropriate rescaling, we compared our estimates
with those of other studies with some clusters in common. We consider
the optical mass estimates by Biviano et al. (1993; 56 clusters), Bird
(1995; 18 clusters), and Girardi et al. (1997a; 8 clusters) finding
that our estimates are smaller by $40\%$, $40\%$, and $20\%$,
respectively.  The difference with Biviano et al.  is partially due to
our more rigorous rejection of interlopers.  The remaining difference
and the difference with Girardi et al. are explainable by our
application of the $C$-correction (due to the surface term in the
virial theorem).

The difference with Bird's (1995) estimates is due to a combination of
the following effects: i) a different rejection of interlopers (our
$\sigma_P$ are smaller by $7\%$ than Bird's values, leading to a
difference in the masses by a factor of $14\%$); ii) the effect of the
different mass estimator; iii) the effect of the $C$-correction.  Bird
(1995) used the projected mass estimator, which assumes that mass
follows galaxy distribution and that galaxy velocities are isotropic,
whereas the virial estimator does not require any assumption about
velocity anisotropies.  In the case of radial or circular
anisotropies, the virial estimator gives, respectively, larger or
smaller mass estimates that the projected estimator (e.g. Perea et
al. 1990).  However, this different behavior is counterbalanced by
the effect of the $C$-correction which reduces virial masses by a
large amount in the case of radial orbits (clusters of type A in
\S~4.5 and 4.6) and by a small amount in the case of isotropic and
circular orbits (clusters of types B and C).  Given these two combined
effects, for both A clusters and B+C clusters we obtain masses which
differ from Bird's estimates by a similar fraction (note that we have
10:5:3 clusters of types A:B:C).

Our estimate of $C$-correction ($19\%$, median value) is in good
agreement with that given by The \& White (1976) for the Coma cluster
($27\%$, the maximum possible correction) and with that recently
suggested by Carlberg et al. (1997a, $18\%$ mean value for CNOC
clusters).  We stress the fact that our estimate of this correction is
only an approximation for real clusters, since it assumes that the
cluster is entirely in dynamical equlibrium and not only up to the
virialization radius. The overall reliability of this mass correction
is supported a posteriori by the fair agreement between the corrected
virial mass and the mass derived from the Jeans equation.  Moreover, a
quantitatively similar correction is obtained by using an alternative
approach which analyzes the effect of galaxy infall and turbulent
galaxy motions around clusters on observational cluster parameters
(Mezzetti et al.  1998).

The agreement between optical and X-ray masses is well explained in
the context of two common assumptions: i) that mass follows the 
galaxy distribution, ii) that clusters are not far from a situation of
dynamical equilibrium, with both gas and galaxies reflecting the same
underlying mass distribution.
 
Both our estimates of virial masses and the verification with the
Jeans equation are based on the assumption that mass follows the 
galaxy distribution, whereas X-ray masses do not require any a priori
assumption about the cluster mass distribution.  Therefore, the
agreement we find between $M_{opt}$ and $M_X$ provides a good support for the
``mass follows galaxies'' assumption.  In particular, in order to look
for possible departures from this law, we also analyze the relation
between the ratio $M_X/M_{opt}$ and $R_X$ ($R_X$ ranging from $0.06$
to $\sim 2 \times R_{vir}$), finding no significant correlation.  Our
results regarding mass distribution are to be added to considerable evidence
coming from optical, X-ray analyses, and gravitational lensing (e.g.,
Carlberg et al. 1997a; Cirimele et al. 1997; Durret et al. 1994;
Narayan \& Bartelmann 1997; Watt et al. 1992).

The fact that gas should be roughly in dynamical equilibrium within
clusters is shown by several works based on numerical simulations
(e.g., Schindler 1996a; Evrard et al. 1996).  As regards galaxies, by using
CNOC cluster data, Carlberg et al. (1997b) have recently found that
both red and blue galaxy populations are in dynamical equilibrium
within the cluster potential.  

Our relation between optical and X-ray mass estimates shows a
large scatter which cannot be explained by the
observational errors (unless they are strongly underestimated). This
fact suggests the presence of an intrinsic scatter, possibly due to some
deviation from the ``mass follows galaxies'' law or from pure
dynamical equilibrium.

Moreover, we do not rule out  that there may be some clusters which are far
from dynamical equilibrium and, indeed, we cannot make any analysis
for our strongly substructured clusters whose mass value is very
uncertain (Tables~7 and ~8). Therefore, our conclusion is that about
$90\%$ of nearby clusters are not far from dynamical equilibrium, in
agreement with the results of a statistical analysis devoted to
cluster substructures in 48 clusters (Girardi et al. 1997a). For
clusters which lie at moderate or large distances, the question is
still open since evidence both in favor of and against dynamical
equilibrium has been reported (e.g., Allen 1997; Allen, Fabian, \& Kneib
1996; Carlberg et
al. 1997a; Girardi et al. 1997b; Loeb \& Mao 1994; Miralda-Escud\'e \&
Babul 1995; Wu 1994).

On the assumptions of cluster dynamical equilibrium, we can further
investigate the relation between galaxies and gas components.  As a
first approximation, one can assume that the hot gas is isothermal,
and that galaxies have a constant velocity dispersion profile and
isotropic orbits. On these assumptions, the Jeans equation gives
$\beta_{spec}(:=\sigma^2_P\mu m_p/kT)
=\beta_{fit,gas}/\beta_{fit,gal}$ where $\beta_{fit}$ gives the slope of
the density distribution in external cluster regions, $T$ is the gas
temperature, $\mu$ is the mean molecular weight and $m_p$ the proton
mass (e.g., Gerbal, Durret, Lachi\'eze-Rey 1994; Bahcall \& Lubin
1994).  Since the gas distribution is more extended than the galaxy
distribution $\beta_{fit,gas}=0.67$ (e.g., Jones \& Forman 1984)
against $\beta_{fit,gal}=0.8$ as found in this work (see also 
Bahcall \& Lubin 1994 and references therein), we expect that 
the gas is ``hotter'' than the galaxies.  

Indeed, by comparing our $\sigma_P$ with $T$ taken from
David et al. (1993) and from White et al. (1997), we obtain that
$\beta_{spec}=0.88\pm0.04$.  Note  that this value refers to 55 clusters
with better optical data according to the requirements of Girardi et
al. (1996), i.e.  with at least 30 member galaxies and excluding strongly
structured clusters.
The fitted weighted relation is $\sigma_P \propto
10^{2.51\pm0.03} \times T^{0.62\pm0.04}$ (see Figure~6)  which is
consistent with that given in Girardi et al. (1996) but no longer consistent
with the model of perfect galaxy/gas equipartition
($\beta_{spec}=1$). This relation suggests some dependence of
$\beta_{spec}$ on cluster mass (see also Bird, Mushotzky, \& Metzler
1995).  

The above phenomena can be explained on the view that galaxies are
slowed down by dynamical friction (e.g. Bird et al. 1995) and/or that
gas is heated by galactic winds (e.g., White 1991) or during a
preheating phase of the intracluster medium (e.g., Cavaliere, Menci,
\& Tozzi 1997).  Recent studies based on numerical simulations, even
when galactic winds are included (e.g., Metzler \& Evrard 1997),
suggest that the effect on the galaxies is the main one, since they
found the presence of a velocity bias, i.e. galaxies slowed down with
respect to dark matter, $DM$.  But in this scenario the $DM$
distribution is more extended than the galaxy distribution, hence
virial masses should strongly underestimate cluster masses (by up to a
factor of 5, according to Frenk et al. 1996). However, these results
on DM distributions are in disagreement with the results of X-ray
analyses (e.g., Cirimele et al. 1997; Durret et al.  1994; Watt et
al. 1992) which, instead, give evidence of a DM distribution less
extended than that of gas as expected in the case of heating of the
gas to temperatures higher than those of galaxies and DM.

\includegraphics{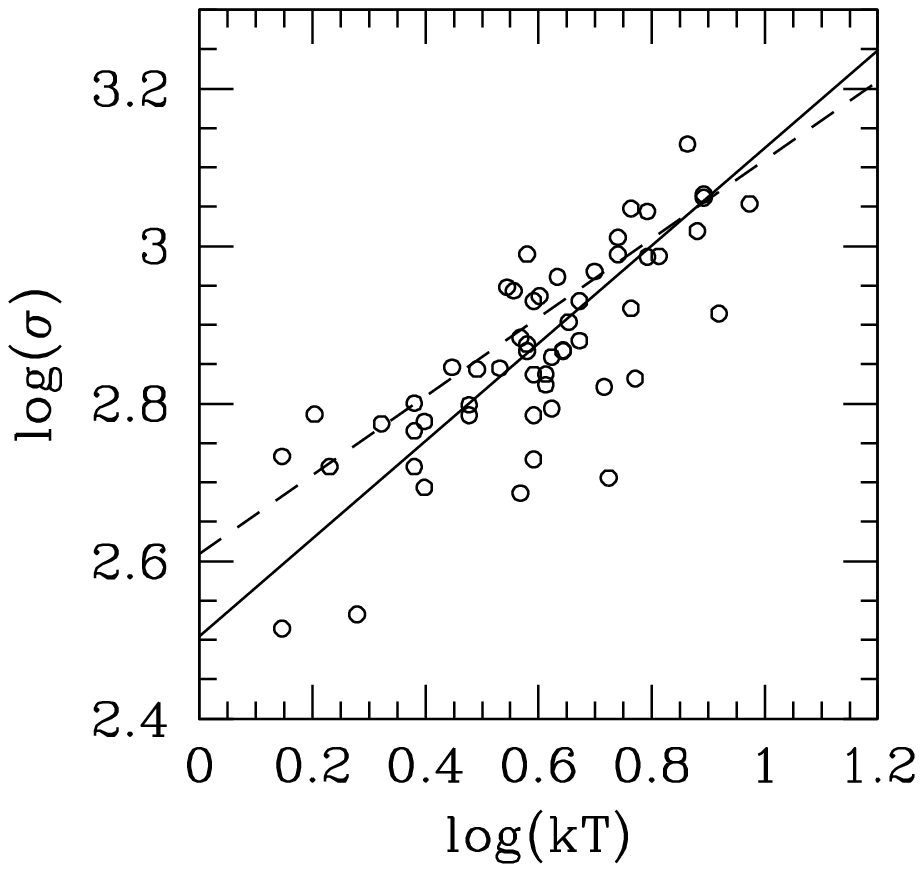}
$\ \ \ \ \ \ $\\
\vspace{6truecm}
$\ \ \ $\\
{\small\parindent=3.5mm {Fig.}~6.---
The solid line is the fit (weighted regression line)
 on 55 clusters. The dashed line represents the model with $\beta_{spec}=1$.
}
\vspace{5mm}

Although our work is not devoted to the analysis of galaxy
distribution, we have had to address this issue in the computation of
cluster masses. By using  a King-like profile we obtain good
alternative estimates of individual virial radii (see Figure~1,
top-left panel), which is our aim in order to obtain reliable cluster
mass estimates. Moreover, when one or two parameters ($\alpha$,$R_c$)
are fixed, the estimate of virial radii result less good, supporting
the existence of an intrinsic spread of cluster parameters.

As for the comparison with previous King-like fits, we find good
agreement with the value of the exponent $\alpha$ obtained by G95 ($\alpha
=0.8^{+0.3}_{-0.1}$) and with other previous estimates (see Bahcall \&
Lubin 1994 and references therein).  Instead, we find typically
smaller core radii $R_c$ (cf. our median $R_c=0.05$ \h~with $R_c=0.17$
\h~ of G95).  We verify that this difference is due to the fact that
we determine cluster centers from the galaxy density peak (by using
the two-dimensional kernel adaptive method, e.g. Girardi et al. 1996;
Pisani 1996) rather than from averaging galaxy coordinates. Our finding
is in agreement with the suggestion of Beers \& Tonry (1986) that
large core radii can be produced by inaccurate cluster
centers. Noticeably our small values for $R_C$ are in agreement with
core radii relative to the total mass distribution derived from
gravitational lensing (upper limits from 30 to 65 \kk, Narayan \&
Bartelmann 1997).

\includegraphics{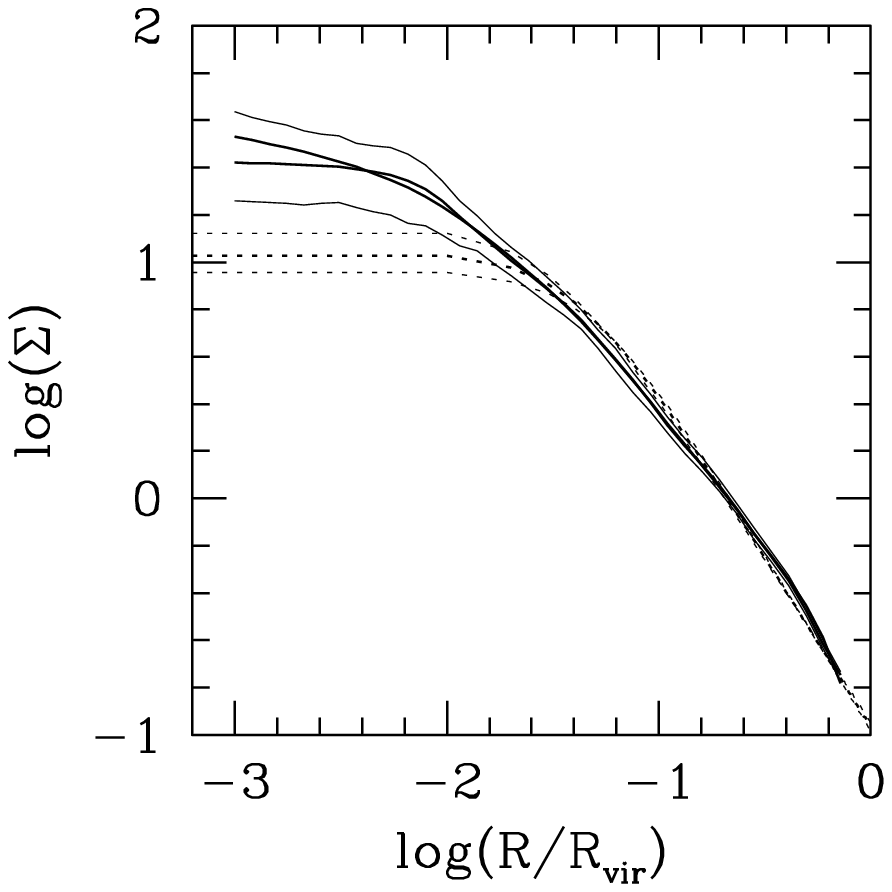}
$\ \ \ \ \ \ $\\
\vspace{6truecm}
$\ \ \ $\\
{\small\parindent=3.5mm {Fig.}~7.---
The comparison of the fits to the (normalized)
combined projected galaxy number density $\Sigma$.  The dotted line
gives the result of the King-like profile fit with its error bands
(faint dotted lines).  The solid lines give the results of the nonparametric
(MPL) fit with two smoothing parameters; the more smoothed solution
showing a stronger gradient (in central regions) 
than the less smoothed solution for which
we also give the bootstrap error bands (at the $95\%$ s.l., faint
solid lines).
}
\vspace{5mm}

However, the question of the existence of a core radius in the galaxy
distribution is still open.  For instance, Merritt \& Tremblay (1994)
find no evidence for a core in Coma clusters (by using about 1500
galaxies), whilst Adami et al.  (1998c) found that King profile better
fits the ENACS clusters than do de Vaucouleurs or Navarro et
al. (1996) profiles (see also Adami et al.  1998b).  Indeed, we use the
King-like profile in order to obtain information for individual
clusters, but, to go deeper in the ``core radius'' issue we use a
nonparametric method to fit the galaxy number density obtained by
combining together the galaxies of all well sampled clusters (92
clusters sampled up to $R_{vir}$, see \S~4.3) and normalizing
distances to $R_{vir}$.  In particular, we use the Maximum Penalized
Likelihood (MPL) method (Merritt \& Tremblay 1994) 
which gives accurate representations of the
true density profile according to the authors.
Figure~7 compares the fits to the surface galaxy number density: the
result of the King-like profile ($R_c=(0.04\pm 0.01) \times R_{vir}$,
$\alpha=0.72\pm 0.03)$, and the MPL fits.  For MPL we show two
estimates derived using two different smoothing parameters
($\lambda=10^{-4},10^{-5}$; see Merritt \& Tremblay 1994 for details), which
represent a compromise in computing a solution which is ``most
consistent'' with the data, but not unacceptably noisy; the real
features are those that persist over a wide range of choices for the
smoothing.  The existence of a core radius, being dependent on the
choice of the smoothing parameter, is questionable.  There is no
evidence for a core radius as large as the one fitted by the King-like
profile, but rather there is an indication for a smaller core radius
($\sim 0.01 \times R_{vir} \sim 10$ \kk) or, alternatively, no core radius.
It is difficult to address this issue since we are dealing with
dimensions close to galaxy sizes.  However, the outer ($R\gtrsim 50$
\kk) profile given by both the MPL estimates is in acceptable
agreement with the profile of the King-like fit corroborating our use
of the King-like profiles in estimating ``global'' quantities.

\section{SUMMARY AND CONCLUSIONS}

The main points of this work may be summarized as follows:

i) We evaluate in a homogeneous way the optical masses of 170 nearby
clusters ($z\le 0.15$).  This sample, which is the largest set of
clusters up to now analyzed in the literature, includes both data from
the literature and the new ENACS data (Katgert et al. 1996, 1998).

ii) On the assumption that mass follows the galaxy distribution, we
compute the masses of each cluster by applying the virial theorem
to the member galaxies and we verify our results by using the Jeans
equation. 

iii) Our mass estimates are smaller than previous optical
estimates. This fact is due both to our better membership assignment
procedure and to the application of the 
correction due to the presence of the surface term in the virial theorem
(recently stressed by Carlberg et al. 1997a).

iv) After appropriate rescaling to the X-ray radii, we compare our
optical mass estimates to those derived from X-ray analyses, which we
have compiled from the literature (for 66 clusters).  We find a good
overall agreement.

v) The above agreement is expected on the basis of two common
assumptions: a) that mass follows the galaxy distribution, b) that
clusters are not far from a situation of dynamical equilibrium with
both gas and galaxies reflecting the same underlying mass
distribution. It should be pointed out that Carlberg et al. (1997a)
have recently drawn similar conclusions for a sample of distant
clusters (the CNOC sample).  In particular, we find evidence for a
galaxy distribution which is colder and less extended than the gas
distribution. 

Several recent studies have casted doubts on cluster mass estimates
and attempted to lower the cluster baryon fraction by reducing the
cluster masses (e.g., Gunn \& Thomas 1996; Wu \& Fang 1996).  We
stress that our study strongly supports the reliability of present
cluster mass estimates derived from X-ray analyses and/or
(appropriate) optical analyses.  Hence, it is even more difficult to
reconcile present data with a $\Omega_0=1$ Universe (e.g. White et
al. 1993b).

Our cluster masses are suitable for statistical studies.  In
particular, we did not reject a priori those clusters with a poor number
of selected members, which usually have a small 
mass, in order to avoid having a final cluster sample biased towards
more massive systems.

\acknowledgments

We are particularly indebted to the ENACS team for having kindly
provided us with their data, based on observations collected at the
European Southern Observatory, in advance of publication.  

We thank Dario Fadda for useful discussions and the anonymous referee
for useful suggestions.  This work has been partially supported by the
Italian Ministry of University, Scientific Technological Research
(MURST), by the Italian Space Agency (ASI), and by the Italian
Research Council (CNR-GNA).

\end{multicols}
\small

\end{document}